\documentclass{elsarticle}

\usepackage{graphics}
\usepackage{amsmath}
\usepackage{amssymb}

\newcommand{\eps}{\epsilon}
\newcommand{\z}{\zeta}
\newcommand{\tends}{\rightarrow}
\newcommand{\Li}{\text{Li}}

\journal{Physica A}

\begin{document}

\begin{frontmatter}

\title{Bose-Einstein condensation in the three-sphere and in the infinite slab: analytical results}

\author[ulpor]{J.M.B.~Noronha}
\ead{jnoronha@por.ulusiada.pt}
\author[ncl]{D.J.~Toms}
\ead{david.toms@newcastle.ac.uk}
\address[ulpor]{Universidade Lus\'{\i}ada -- Porto, R. Dr. Lopo de Carvalho, 4369-006 Porto, Portugal}
\address[ncl]{School of Mathematics \& Statistics, Newcastle University,
Newcastle upon Tyne, NE1 7RU, U. K.}

\begin{abstract}
We study the finite size effects on Bose-Einstein condensation (BEC) of an ideal non-relativistic Bose gas in the three-sphere (spatial section of the Einstein universe) and in a partially finite box which is infinite in two of the spatial directions (infinite slab). Using the framework of grand-canonical statistics, we consider the number of particles, the condensate fraction and the specific heat. After obtaining asymptotic expansions for large system size, which are valid throughout the BEC regime, we describe analytically how the thermodynamic limit behaviour is approached. In particular, in the critical region of the BEC transition, we express the chemical potential and the specific heat as simple explicit functions of the temperature, highlighting the effects of finite size. These effects are seen to be different for the two different geometries. We also consider the Bose gas in a one-dimensional box, a system which does not possess BEC in the sense of a phase transition even in the infinite volume limit.
\end{abstract}


\end{frontmatter}

\section{Introduction}
\label{intro}

The phenomenon of Bose-Einstein condensation (BEC) is described in statistical mechanics text-books (e.g. \cite{Huang,Pathria}). Given an ideal gas of particles obeying Bose statistics inside a box with sides of length $L_1$, $L_2$, $L_3$, as the thermodynamic limit is taken in the usual way ($N\tends\infty$, $V\tends \infty$ while $N/V$ and $L_i/L_j$ are held fixed), the fraction of particles in any excited state $i$,  $N_i/N$, goes to zero. This is expected since the single particle energy levels get closer and closer to each other, eventually forming a continuum in the thermodynamic limit. The ground state is the exception. Indeed, below a certain critical temperature, the fraction of particles in the ground state, $N_{\text{gr}}/N$, will be non-vanishing in the thermodynamic limit, which means that the probability distribution of particles as a function of their energy has a Dirac $\delta$ component at zero energy. Because of this, the chemical potential and, as a consequence thermodynamic functions in general, are non-analytical at the critical temperature, signalling the BEC phase transition.

In finite systems, however, we are away from the thermodynamic limit and, strictly speaking, a phase transition does not happen, all thermodynamic functions being smooth functions of the temperature at the critical point. Nevertheless, a large finite system can be practically indistinguishable from an infinite one. For example, in the thermodynamic limit as described above, the specific heat for the three-dimensional gas has a sharp peak with a discontinuous first derivative at the onset of BEC. For a finite large system, there will still be a more or less sharp peak but all thermodynamic functions will be analytical. The larger the volume that the finite system has, the sharper the peak. The difference between a finite system and its infinite counterpart can be brought out analytically in the form of finite size correction terms. Furthermore, these finite size corrections are dependent on the particular way the gas is confined; namely, they depend on the geometry of the system.

It is our aim in this article to obtain analytically finite size corrections to the thermodynamics of an ideal non-relativistic Bose gas in the quantum degenerate regime in three distinct physical situations: the three-sphere (spatial section of the Einstein universe), the three-dimensional box which is infinite in two of the directions and finite in the other one (we will call this the infinite slab) and the one-dimensional box.
The choice of these models enables us to perform a full analytical treatment showing very explicitly in simple expressions the finite size effects and the impact that different geometries can have in these effects.
The three seemingly disparate situations under consideration
have mathematical aspects in common, allowing for a unified treatment as will be seen. Actually, it is well known that the one-dimensional box does not possess BEC, but once we have all the mathematical apparatus set up, the study of this system requires no extra effort. The infinite slab is an example of generalized BEC, in which the particles condense into a low lying set of states rather than the ground state, as described early on in \cite{Sonin} 
(and which the work in \cite{Krueger} already hinted at) 
and later systematically studied by van den Berg and collaborators \cite{vandenBerg1,vandenBerg2,vandenBerg3} (see \cite{BeauZagrebnov,Mullin} for recent articles on this topic). From the systems we study here, the only one having the usual form of BEC is the three-sphere.
 
Our point of departure in the analytical treatment of the thermodynamic sums will be the Mellin-Barnes transform, a tool used in the past in \cite{KT:HO,StandenToms98,StandenToms99,Toms06}, to study respectively the ideal Bose gas in a harmonic oscillator potential, in the infinite flat space subject to a magnetic field and in the three dimensional space in which one of the directions is compactified to form a circle. 
This transform can also be used to obtain high temperature expansions for the ideal Bose gas in quite general settings, as was shown for the Bose gas under arbitrary background potentials and in boxes of arbitrary shape \cite{KT98,KT99} and, more recently, for the Bose gas in product manifolds \cite{FucciKirsten}. Although these high temperature expansions can provide a way of calculating a BEC critical temperature in each setting, they cannot be used to study the BEC transition itself or the BEC regime.
Our approach allows us  to obtain large size expansions, in terms of temperature and chemical potential,
 which are valid and very effective throughout the quantum degenerate regime and in the vicinity of the critical region. This is done in section~\ref{NC}. For the first two systems mentioned above, we will then obtain, in section~\ref{critical}, an analytical description of the approach to the critical behaviour which signals the onset of BEC in the thermodynamic limit. Specifically, in the critical region we obtain the chemical potential, fraction of condensed particles and the specific heat as explicit functions of the temperature only. From the results thus obtained, it will be clear that the geometry of the system has a crucial role in the nature of the finite size corrections in each situation. For example, while in the three-sphere the specific heat peak happens at a slightly higher temperature than the temperature at which the chemical potential is zero, the opposite happens in the infinite slab scenario.

The case of the one-dimensional box has been treated before by Pathria \cite{Pathria98} who obtained finite size corrections to the number of particles (as a function of temperature and chemical potential) using the Poisson summation formula. However, his procedure is valid only away from the quantum degenerate regime. Finite size corrections to non-relativistic BEC in the three-sphere have also been treated before by Altaie \cite{Altaie78}. His procedure is similar to that of Pathria's article mentioned above. In particular, it is based on the use of the Poisson summation formula and likewise, it is valid only away from the quantum degenerate regime. A physical system analogous to our infinite slab case was considered before in the context of the thermal Casimir effect in \cite{Martin,Biswas}, where the authors obtain high temperature (classical region) expansions for the thermodynamic potential and tackle the quantum degenerate regime by setting the chemical potential identically equal to its (thermodynamic limit) critical value.
Some developments were also made in the finite size effects of a relativistic Bose gas in the three-sphere \cite{Trucks,AltaieMalkawi}. These works used a combination of analytical and numerical techniques and were very much inspired by an earlier study by Singh and Pathria \cite{SinghPathria}. Other work on BEC for the relativistic Bose gas on the 3-sphere includes \cite{SmithToms}. Some work on finite size effects for the uncharged relativistic Bose gas on the 3-sphere have been considered \cite{ZhukKleinert}.

\section{Number of particles and specific heat}
\label{NC}

Consider an ideal Bose gas with particle
eigenstates of energy $E_{n}$. The grand-canonical average number of particles is
     \begin{equation}
    N=\sum_{i} [e^{\beta (E_{i}-\mu )}-1]^{-1}  ,
    \label{Ngeneral}
     \end{equation}
where $\beta $ is the inverse of the temperature $T$, $\mu $ is
the chemical potential and the sum is over all particle
eigenstates. We will use the natural units system with $\hbar =1$, $c=1$, and Boltzmann's constant $k=1$ throughout. Expression (\ref{Ngeneral}) gives us $\mu (T)$ implicitly if we fix
$N$. The internal energy is
     \begin{equation}
    U=\sum_{i} E_i[e^{\beta (E_{i}-\mu )}-1]^{-1}  .
    \label{Ugeneral}
     \end{equation}
The specific heat at constant volume is obtained by differentiating $U$ with respect to $T$ with the volume and number of particles held fixed: $\displaystyle{C=\left.(\partial U/\partial T)\right|_{N,V}}$. In this section, we shall aim at obtaining analytical
expressions for $N$ and $C$ as functions of
$\mu $ and $\beta $, in the form of expansions that contain the thermodynamic limit behaviour plus finite size corrections.

\subsection{The three-sphere}
\label{NC:S3}

The energy eigenvalues in the three sphere are given by
     \[
    E_{n}=\frac{1}{a}\left( n^{2}+m^{2}a^{2}+6\xi -1
    \right) ^{1/2}  ,
    \; \; n=1,2,3,\ldots
     \]
with degeneracy $g_{n}=n^{2}$ (see e.g. \cite{Ford75,Ford76}). $a$ is the radius of the three-sphere,
$m$ is the particle mass and $\xi $ is a coupling constant that describes the interaction of the field with the scalar curvature. We
will limit ourselves to conformal coupling, $\xi =1/6$ for simplicity. As noted
before \cite{AltaieMalkawi}, this is not very restrictive since choosing a
different coupling is equivalent to changing the mass from $m$ to
$[m^{2}+ (6\xi -1)/a^{2}]^{1/2}$. For large $a$, this change is
very small. In particular, in the large volume limit, all
couplings will give the same results. 
In the non-relativistic
limit, $E_n$ can be expanded in powers of $n^2/(ma)^2$ yielding
	\begin{equation}
	E_n=m+\frac{n^2}{2ma^2}\left( 1-\frac{1}{4}\left( \frac{n}{ma}\right)^2+O\left(\left(\frac{n}{ma}\right)^4\right)\right) \; .
	\label{levelsS3}
	\end{equation}
The constant term, $m$, can be absorbed into the chemical potential in (\ref{Ngeneral}) and (\ref{Ugeneral}), which amounts in fact to a redefinition of the chemical potential (from $\mu$ to $\mu -m$) in all expressions, with no effects in the physics of the system. Hence, we will drop $m$ from (\ref{levelsS3}). In addition, we will consider only the first term inside the parenthesis, leaving only $E_{n}\simeq n^{2}/(2ma^{2})$.
Taking this limit requires $n\ll ma$, i.e., little occupation of
high energy states (``high'' when compared with $m$),
corresponding to $T\ll m$. For this to be true in the
temperature range we are most interested in, the critical region,
it suffices having $\rho \ll m^{3}$, where $\rho $ is the particle
density. 
The relative error involved in this non-relativistic approximation is of the order $T/m$.\footnote{ To see this, we split the sums in (\ref{Ngeneral}) and (\ref{Ugeneral}) into a non-relativistic main part running from $n=1$ to $n=p$, plus a tail (where the non-relativistic approximation is not valid) running from $n=p+1$ to infinity. $p$ must be a number large enough so that the contribution of the tail to the whole sum can be neglected and, at the same time, small enough for the non-relativistic approximation to be valid in the first part of the sum. By choosing $p=ma(T/m)^r=(2x)^{-1/2}(T/m)^{r-1/2}$, with $1/4<r<1/2$, it can be shown that for small $T/m$ and small $x$, the relative error in the first part of the sum is of the order $T/m$ and that the tail contribution to the error is exponentially suppressed. (See also next footnote.)}

We have then
     \begin{equation}
    N=\sum_{n=1}^{\infty } n^{2}\left[e^{x(n^{2}+\eps
    )}-1\right]^{-1} 
    =\sum_{n=1}^{\infty }
    \sum_{k=1}^{\infty }n^2e^{-kx(n^{2}+\eps )}  ,
    \label{NS3}
     \end{equation}
where we have changed to the dimensionless variables,
     \begin{equation}
    x=\frac{\beta }{2ma^{2}}  \, , \qquad
    \eps =-2ma^{2}\mu \, .
    \label{x,eps}
     \end{equation}
(Note that $\eps >-1$ always, for positiveness of the occupation
numbers, and that $\eps \rightarrow -1$ implies $N\rightarrow
\infty $.) $x$ and $\eps$ can be seen as the dimensionless inverse temperature and chemical potential respectively. The number of particles in the ground state is $N_{\text{gr}}=\left[ e^{(1+\eps)x}-1\right] ^{-1}$ when expressed in terms of dimensionless variables.

In the thermodynamic limit we have $x\tends 0$, which yields the bulk result
     \begin{equation}
    N_{\text{bulk}}=\frac{\sqrt{\pi}}{4}\Li_{3/2}(e^{-x\eps})x^{-3/2}
    \label{Nbulk1-S3}
     \end{equation}
when $\eps\geq 0$ and
     \begin{equation}
    N_{\text{bulk}}=\frac{1}{x(1+\eps)}+\frac{\sqrt{\pi}}{4}\z\left(
    \frac{3}{2}\right) x^{-3/2}
    \label{Nbulk2-S3}
     \end{equation}
when $-1<\eps<0$.\footnote{The first order relativistic corrections to expressions (\ref{Nbulk1-S3}) and (\ref{Nbulk2-S3}) for small $T/m$ are respectively $(15\sqrt{\pi}/32)(T/m)\text{Li}_{5/2}(e^{-x\epsilon})x^{-3/2}$ and $(15\sqrt{\pi}/32)(T/m)\zeta (5/2)x^{-3/2}$. (Hence, the relative error involved in the non-relativistic approximation is of order $T/m$, as previously stated.)}
 $\Li_{\alpha}(z)=\sum_{n=1}^{\infty}z^n/n^{\alpha}$ denotes the polylogarithm
function. In
(\ref{Nbulk2-S3}), the first term on the right hand side is the
ground state contribution (when $(\eps+1)x\ll 1$) and it is of
leading order only if $(1+\eps)\propto x^{1/2}$. Since
the volume of the three-sphere is given by $V=2\pi^2a^3$, we have
$Nx^{3/2}=(\beta/(2m))^{3/2}2\pi^2\rho$, where $\rho$ is the
particle density. Hence, when taking the thermodynamic limit,
$Nx^{3/2}$ is a constant. Since $\Li_{3/2}(e^{-\eps x})$ is a
decreasing function of $\eps x$, defined only when $\eps x\geq 0$, the maximum value that it can
take happens when $\eps=0$ and it is given by $\Li_{3/2}(1)=\z(3/2)$.
Hence, if the particle density exceeds a certain critical value,
$\rho_c$, given by
     \begin{equation}
    \left(\frac{\beta}{2m}\right)^{3/2}2\pi^2\rho_c=\frac{\sqrt{\pi}}{4}
    \z\left(\frac{3}{2}\right)  ,
    \label{rhocrit-S3}
     \end{equation}
$\eps$ must go to negative values (it must go close enough to -1)
allowing the ground state to take the particles that cannot be
accommodated in the excited states and BEC will happen. Note that this $\rho_c$
is the same as the one for the standard case of an infinite cubic
box \cite{Huang,Pathria}. This is natural, since in the thermodynamic limit we have
$a\tends \infty$ and the three-sphere will resemble more and more
an infinite flat space. In fact, by writing $x$ in (\ref{Nbulk1-S3}) and (\ref{Nbulk2-S3}) in terms of volume and temperature, we see that these expressions are exactly the same as those of the infinite cubic box.

From the critical density, given as a function of the temperature
in (\ref{rhocrit-S3}), we have the familiar critical temperature given
as a function of the particle density as $T_c=(2\pi/m)(\rho/\z(3/2))^{2/3}$.
Equivalently, the critical value of $x$ is given by $Nx_c^{3/2}=
\sqrt{\pi}\z(3/2)/4$, as can be seen directly from
(\ref{Nbulk1-S3}).

For $\eps\geq 0$, from (\ref{Nbulk1-S3}) we have that in the
bulk, $\eps x$ is a function of the rescaled temperature, $T/T_c$,
only (equivalently, $x_c/x$). More specifically, it is an increasing function of $T/T_c$,
as can be clearly seen from (\ref{Nbulk1-S3}). Hence, away from the condensate region (i.e.,
$\eps>0$), as we approach the thermodynamic limit, we have
$\eps\propto x^{-1}$ (for fixed $T/T_c$). In the condensate regime
and fixed $T/T_c$ we will have $\eps\tends -1$ in such a way that
$(\eps+1)\propto\sqrt{x}$, so that the condensate density is not
vanishing .

Of course, these results are only correct in the thermodynamic
limit. We now obtain finite size
corrections to the number of particles in (\ref{Nbulk1-S3}) and
(\ref{Nbulk2-S3}) and to the specific heat sums, still to be
introduced.

In \ref{sumsSij}, we define a class of sums, $S(i,j)$, which appear throughout the cases we study here, both in the number of particles and in the specific heat. For these sums we have obtained asymptotic expansions for small $|\eps x|$. These expansions are given in (\ref{A:asympt1}) and (\ref{A:asympt2}). Thus,
from (\ref{NS3}), (\ref{A:def-S}) and (\ref{A:asympt1}) we have 
 \[
N=S(1,0)\sim \frac{\sqrt{\pi }}{4}x^{-3/2}
    \sum_{k=0}^{\infty }\frac{(-1)^{k}}{k!}\z \left(\frac{3}{2}-k\right)
    (\eps x)^{k}
    +x^{-1}f_{1}(1,\eps ) \, , 
      \]
where $f_1(\alpha ,\eps)$ is a function defined in (\ref{A:def-f}). The symbol $\sim$ has the usual meaning of asymptotic expansion. Specifically, by truncating the infinite sum above at a certain order $k=n$, we know that the error in $N$ (or in any other of the sums $S(i,j)$) will be of order $\text{o}\left( (\eps x)^n\right)$.
From the earlier discussion based on the bulk
expressions (\ref{Nbulk1-S3}) and (\ref{Nbulk2-S3}), we know
that for large $N$, the condition $|\eps x|\ll 1$ is satisfied throughout the quantum degenerate regime and also for temperatures not much
higher than $T_c$. Therefore, for these temperatures, the number
of particles should be well approximated by only a few terms from
the sum in $k$ above plus the isolated term. The functions $f_i(\alpha ,\eps)$ are studied in \ref{functionsfi}, where we find their meromorphic structure and their values at particular points. We finally obtain
     \begin{equation}
    N\sim \frac{\sqrt{\pi}}{4}x^{-3/2}
    \left[ \z \left(\frac{3}{2}\right)
    -\z \left(\frac{1}{2}\right)\eps x-\frac{1}{8\pi }
    \z \left(\frac{3}{2}\right)(\eps x)^{2}\right] 
    -\frac{\pi \sqrt{\eps }}{2}\coth (\pi \sqrt{\eps })x^{-1} , 
    \label{Ntrunc}
	\end{equation}
where we have truncated the sum at the third term and for the last term we require the value of $f_1(1,\eps)$ from (\ref{B:recur}), (\ref{B:f0at1}) and (\ref{B:f0at-k}). Note that this expression is valid for both positive and negative $\eps$, although for $\eps <0$  the last term is more conveniently written with a regular cotangent by noting that 
$\sqrt{\eps }\coth (\pi \sqrt{\eps })=\sqrt{-\eps }\cot (\pi \sqrt{-\eps })$.

As mentioned in the introduction, the non-relativistic ideal Bose gas in the three-sphere has been treated before by Altaie \cite{Altaie78}. In particular, using a procedure which is valid in the $\eps >0$ region only (i.e., away from the quantum degenerate regime), 
Altaie obtained an approximate expression
for $N$ which consists of the first term inside the square brackets and the term outside the square brackets in (\ref{Ntrunc}), which amounts to truncating the asymptotic series for $N$ above at the $k=0$ term. Hence, our result shows that this is simply the leading part of a full asymptotic expansion which is valid for both positive and negative $\epsilon$ (i.e., also in the BEC regime, where in fact it is most useful) and which can be made as accurate as one wishes (at least, in a temperature window around $T_c$) by truncating at an appropriate order.

When we take the thermodynamic limit we have $x\rightarrow 0$ and
we recover the continuum approximation given by
(\ref{Nbulk1-S3}) and (\ref{Nbulk2-S3}), as would be
expected. To see this, note that the expansion in square brackets
is the small $\eps x$ expansion of $\Li_{3/2}(e^{-\eps x})$ obtained by Robinson in \cite{Robinson} except that the term
$-2\sqrt{\pi}(\eps x)^{1/2}$, present in the polylogarithm expansion, is not present here. However, when $\eps\tends \infty$
(which happens for $T>T_c$ when $N\tends\infty$) the hyperbolic
cotangent tends to 1 exponentially fast. Hence, the last term in
(\ref{Ntrunc}) will account for the missing polylogarithm
expansion term and $N$ will be given by (\ref{Nbulk1-S3}). On
the other hand, in the BEC regime we will have $\eps\tends -1$
when $N\tends\infty$ and, expanding
$\sqrt{\eps}\coth(\pi\sqrt{\eps})$ in powers of $(\eps+1)$, the
last term in (\ref{Ntrunc}) can be written as
 \[
     -\frac{\pi \sqrt{\eps }}{2}\coth (\pi \sqrt{\eps })x^{-1}=
    x^{-1}\left[\frac{1}{\eps+1}-\frac{3}{4}-\left(
    \frac{\pi^2}{12}+\frac{1}{16}\right) (\eps+1)+\cdots\right] .
	\]
Thus, to first order and when $\eps\simeq -1$, $N$ is given by
(\ref{Nbulk2-S3}). The -3/4 term in this expression gives us
the first order finite size correction to the number of particles.

We now turn our attention to the specific heat. In the case of
non-relativistic particles in the three-sphere, by using (\ref{Ugeneral}),
(\ref{x,eps}) and the energy levels $E_{n}= n^{2}/(2ma^{2})$ with degeneracy $g_n=n^2$, we have for the internal energy
     \begin{equation}
    U=\frac{1}{2ma^{2}}\sum_{n=1}^{\infty }
    n^{4} \left[ e^{x(n^{2}+\eps )}-1\right] ^{-1}  .
    \label{U-S3}
     \end{equation}
For the specific heat at constant radius $a$, we obtain
     \begin{equation}
    C=x^{2}
    \left[ S(3,1)-\frac{S(2,1)^{2}}{S(1,1)}\right]  ,
    \label{sh-S3}
     \end{equation}
where we used $\displaystyle{\left.\left( \partial
N/\partial x\right)\right|_{N,a}=0}$ to obtain $\displaystyle{\left.\left( \partial
(x\eps )/\partial x\right)\right|_{N,a}}$. Inside the square brackets we have the omnipresent sums $S(i,j)$ defined in (\ref{A:def-S}), which once again have asymptotic expansions for small $\eps x$ given in (\ref{A:asympt1}).
Altaie also obtained expressions for the specific heat sums, but these are only valid in the $\epsilon >0$ region as they yield an imaginary specific heat when $\epsilon <0$.

In figure~1,
we show the fraction of particles in the ground state, $N_{\text{gr}}/N$, as a function of the rescaled temperature, $T/T_c$. In figure~2 we show 
the specific heat per particle. For both plots we used (\ref{Ntrunc}) (with only the terms explicitly displayed) to obtain the values of $\eps$ to input in $N_{\text{gr}}$  (for figure~1) and in the sums that appear in (\ref{sh-S3}) (for figure~2). For these sums, we used their asymptotic expansions (\ref{A:asympt1}) truncating the $k$ sums at $k=2$. In both figures, we can clearly see the approach to the familiar thermodynamic limit critical behaviour as $N$ is increased. In the next section, we provide an analytical description of this phenomenon. 
Note that contrary to what happens in the case of a Bose gas in a harmonic trap, in our case the finite size effects bring the features which are characteristic of the BEC transition to higher values of the temperature. This also happens in the case of a Bose gas in a box \cite{KT99,Grossmann} or in a power-law trap with power higher than 3 \cite{Jaouadi}.
Had we
also plotted the exact numerical results given by (\ref{NS3}), (\ref{sh-S3}) and (\ref{A:def-S})
the figures would not look any different because the numerical plots
would totally superimpose the analytical ones. 
In fact, our analytical approximation is extremely good for temperatures near $T_c$ and throughout most of the condensate region. Taking as an example the $N=10^2$ case, the relative error in the condensate fraction, $(N_{\text{gr}}/N(\text{analytical})-N_{\text{gr}}/N(\text{numerical}))/(N_{\text{gr}}/N(\text{numerical}))$, is never more than $3\times 10^{-6}$ for $0.15<T/T_c<1.3$ being of the order of $10^{-8}$ at $T=T_c$. For temperatures lower than $\sim 0.15 T_c$ it deteriorates due to the increase in $x$ ($x\rightarrow\infty$ as $T\rightarrow 0$), while $\epsilon\simeq -1$. As the temperature rises above $T_c$, the approximation deteriorates due to the increase in $\epsilon x$. At $T=2T_c$ or $T=0.01T_c$ the relative error is already $6\times 10^{-4}$ (still quite small). For $N=10^5$ the error is similar in the classical regime (since $\epsilon x$ is largely insensitive to $N$ in this regime) but much smaller in the condensate regime (of the order of $10^{-10}$ or less in the critical region and in most of the condensate regime). The relative error in the specific heat has a similar behaviour, although it is somewhat larger (very roughly, by a factor of about 10 in most regions). Even higher precision is easily attained by truncating the asymptotic series at a higher order. Less precision is attained if we truncate at a lower order. Had we truncated the asymptotic expansion for $N$ at $k=0$, we would get a relative error in the condensate fraction of order $0.1$ already for $T=1.2T_c$, independently of $N$. In the condensate region, it would be a few percent (order $10^{-2}$) in the $N=10^2$ case and of order $10^{-3}$ in the $N=10^5$ case.

\subsection{The one-dimensional box}
\label{NC:1Dbox}

We now consider an ideal Bose gas in a one-dimensional box with
Dirichlet boundary conditions. In this case, the non-relativistic
energy levels are
     \[
    E_{n}=\frac{\pi ^{2}}{2mL^{2}}n^{2} ,\; \; n=1,2,3\ldots
     \]
where $L$ is the length of the box. The levels are non-degenerate. After the change of variables,
     \[
    x=\frac{\beta \pi ^{2}}{2mL^{2}}  \, , \qquad
    \eps=-\frac{2mL^{2}}{\pi ^{2}}\mu  \, ,
     \]
and using (\ref{Ngeneral}), we have
     \begin{equation}
    N=\sum_{n=1}^{\infty } \left[ e^{x(n^{2}+\eps
    )}-1\right]^{-1} 
     =  S(0,0)  \, .
    \label{N1Dbox}
     \end{equation}
Looking at this expression, the only thing that distinguishes this
gas from the one in the three-sphere is the levels
degeneracy.

In the thermodynamic limit, we have the
bulk expression
     \begin{equation}
    N_{\text{bulk}}=\frac{\sqrt{\pi}}{2}x^{-1/2}\Li_{1/2}(e^{-\eps x})  \, ,
    \label{Nbulk-1D}
     \end{equation}
which is valid to first order. In this case, we do not need to
take special care about the ground state because, unlike the
polylogarithm of index 3/2 of the previous case,
$\Li_{1/2}(e^{-\eps x})$ is unbounded when $\eps x\tends 0$. This
means that, independently of the value we set the temperature and
density at (or equivalently, the value we set $Nx^{1/2}$ at), all
particles can be accommodated in a smooth distribution over the
energy levels, without any condensation. BEC does not happen. Another
thing that can be seen from (\ref{Nbulk-1D}) is that, in the
bulk, $\eps x$ is a well defined function of $Nx^{1/2}$ (or in
more physical variables, a function of $\rho/\sqrt{T}$), similarly to the three-sphere case.
In the present case, however, this holds for all
temperatures.

Using (\ref{A:asympt2}) for $S(0,0)$ in (\ref{N1Dbox}) we have the asymptotic expansion for small $\eps x$
     \begin{eqnarray}
    N &\sim &\frac{\sqrt{\pi}}{2}x^{-1/2}
    \left[ \z \left(\frac{1}{2}\right)+\frac{1}{4\pi }
    \z \left(\frac{3}{2}\right)\eps x-
    \frac{3}{32\pi ^{2}}\z \left(\frac{5}{2}\right)(\eps x)^{2}\right]
    \nonumber\\*
    &&\mbox{}+\frac{1}{4}-\frac{1}{24}\eps x
    +x^{-1}\left[ \frac{\pi }{2\sqrt{\eps }}\coth (\pi \sqrt{\eps})
    -\frac{1}{2\eps }\right] ,
    \label{Ntrunc1Dbox}
     \end{eqnarray}
where we have truncated both infinite summations in (\ref{A:asympt2}) at $k=2$.

A very similar expression, which is equivalent to this one, was obtained before by Pathria \cite{Pathria98}, making use of the Poisson summation formula in a procedure that is valid only for $\eps >0$. Once again, our method extends a previously found formula to the $\eps <0$ region. 
    
Similarly to the first case studied, to first order, this
expression just gives the bulk result (\ref{Nbulk-1D}). Indeed,
the expansion inside the first square brackets is the small $\eps
x$ expansion of $\Li_{1/2}(e^{-\eps x})$ except for the missing
term $\sqrt{\pi}(\eps x)^{-1/2}$ (which is the term that provides
the correct divergent behaviour of the polylogarithm when $\eps
x\tends 0$). This missing term is accounted for by the first term
inside the second square brackets ($\coth (\pi\sqrt{\eps})$ tends
to 1 exponentially fast when $\eps\tends\infty$). Hence, we
recover (\ref{Nbulk-1D}). The other terms in (\ref{Ntrunc1Dbox})
are of second order, i.e., they are finite size corrections.

Because (\ref{Ntrunc1Dbox}) is asymptotic for small $\eps x$ and
$\eps x$ increases with temperature, our analytical results for
the one-dimensional box are valid only in the low temperature (or
high density) regime. The range of validity is increased by
including terms of higher order in $\eps x$.

For the specific heat, we can easily derive
     \begin{equation}
    C=x^{2}\left[ S(2,1)-\frac{S(1,1)^{2}}{S(0,1)}\right]  .
    \label{sh-1D}
     \end{equation}
The behaviour of the specific heat and ground state population is
uninteresting in the 1D box case since all quantities are always
smooth. For this reason, we do not include any plots here.

\subsection{The infinite slab}
\label{NC:slab}

In this section, we consider a Bose gas contained in the space
limited by two infinite parallel planes. By this, we mean a gas contained in
a rectangular box with sides of length $L_1$, $L_2$ and $L_3$, in
the limit of $L_2,L_3\tends\infty$, while $L_1$ and $L_2/L_3$ are
held fixed.

The energy eigenvalues of a non-relativistic particle in such a
box, with Dirichlet boundary conditions, are
     \begin{equation}
    E_{n}=\frac{\pi ^{2}}{2m}\left[
    \left( \frac{n_{1}}{L_{1}}\right) ^{2}+
    \left( \frac{n_{2}}{L_{2}}\right) ^{2}+
    \left( \frac{n_{3}}{L_{3}}\right) ^{2} \right]  , \,
    n_{i}=1,2,3\ldots
    \label{levels-slab}
     \end{equation}
Inserting (\ref{levels-slab}) in (\ref{Ngeneral}), we have
     \begin{equation}
    N=\sum_{k=1}^{\infty } \sum_{n_{1}=1}^{\infty }
    e^{-k\beta \left( \frac{\pi ^{2}n_{1}^{2}}{2mL_{1}^{2}}-\mu
    \right) }
    \sum_{n_{2}=1}^{\infty }
    e^{-k\beta \frac{\pi ^{2}n_{2}^{2}}{2mL_{2}^{2}} }
    \sum_{n_{3}=1}^{\infty }
    e^{-k\beta \frac{\pi ^{2}n_{3}^{2}}{2mL_{3}^{2}}}.
    \label{NTaylor-R2}
     \end{equation}
Now, in the
limit $L_{2},L_{3}\rightarrow \infty $ the sums in $n_{2}$ and
$n_{3}$ become integrals.\footnote{Strictly speaking, the limit operations $L_2,L_3\rightarrow\infty$ must be outside all sums in (\protect\ref{NTaylor-R2}). That they can be performed inside the $k$ and $n_1$ sums is assured by the uniform convergence of the summand in this limit.} The number of particles will also be infinite in this limit. 
It is then convenient to define a new quantity, $\eta $,
which will be useful later, as $\eta =NL_1^2/(L_2L_3)$. 
$\eta $ is the number of particles in a cube with sides of length
$L_{1}$. It is finite (as long as $L_1$ is finite)  and we prefer using it instead of the particle density,
as $\eta $ is more suitable as a mathematical analogue of $N$ in
the previous cases.

Performing the integrals in $n_2$ and $n_3$ and changing to the dimensionless variables   
	 \[
    x=\frac{\beta \pi ^{2}}{2mL_{1}^{2}} \, , \qquad
    \eps =-\frac{2mL_{1}^{2}}{\pi ^{2}}\mu \, ,
     	\]
we obtain
	\begin{equation}
    \eta =-\frac{\pi }{4}x^{-1}
    \sum_{n=1}^{\infty}\ln \left[1-e^{-x(n^2+\eps)}\right]
    = \frac{\pi }{4}x^{-1}S(0,-1)\, .
    \label{eta}
	\end{equation}
Again, one of the sums $S(i,j)$ defined in (\ref{A:def-S}) makes its appearance.
    
The thermodynamic limit is achieved by taking $L_1\tends \infty$ (with density and temperature kept constant). In this limit, we merely replace the sum in $n$ by an integral, obtaining the bulk quantity
     \begin{equation}
    \eta_{\text{bulk}}=\frac{\pi^{3/2}}{8}\Li_{3/2}(e^{-\eps x})x^{-3/2}  .
    \label{etabulk1}
     \end{equation}
Similarly to the previous cases, for finite $L_1$ this expression gives us the leading behaviour for small $x$ (large $L_1$) when $\eps >0$. 
The situation is identical to the one in the
three-sphere, as can be seen by comparing
(\ref{etabulk1}) with (\ref{Nbulk1-S3}). The maximum value
that the right hand side can take is $\z\left( 3/2\right)(\pi/(4x))^{3/2}$, when $\eps =0$. As we take the thermodynamic limit,
$\eta x^{3/2}=\rho(\beta\pi^2/(2m))^{3/2}$ is a constant. If this
constant is set at a value higher than $(\pi/4)^{3/2}\z\left( 3/2\right)$,
then $\eps$ will have to go to negative values and close enough to
-1. In this case, the $n=1$ part of the sum in (\ref{eta}),
which we denote by $\eta_1$ and is given by $\eta_1=-(\pi /4)x^{-1}\ln \left[ 1-e^{-x(1+\eps )}
    \right]$,
has to be taken into account separately, just like in the three-sphere case. This yields in the thermodynamic limit,
     \begin{equation}
    \eta_{\text{bulk}}=\frac{\pi^{3/2}}{8}x^{-3/2}\z\left(\frac{3}{2}\right)
    -\frac{\pi}{4}x^{-1}\ln (\eps+1)  \, ,
    \label{etabulk2}
     \end{equation}
when $\eps$ is close to -1. The last term is of first order if
$\ln(\eps+1)\propto -x^{-1/2}$. Note that, unlike in S$^3$,
this term is not the ground state contribution.
Instead, it represents the $\eta_1$ contribution to $\eta$ (i.e.,
the contribution of the particles that are not excited in the
$L_1$ direction).

If we set the temperature at a value lower than the critical
value, $T_c$ (which is the same as the one in the three-sphere
 or the one in the more standard cubic box case), the
fraction of particles with $n_1=1$, $\eta_1/\eta$, will be
non-vanishing in the thermodynamic limit. What happens is a generalized BEC \cite{vandenBerg1,vandenBerg2,vandenBerg3}. The
usual form of BEC does not happen at any temperature or density.
Indeed, the ground state occupation number is given by
$N_{\text{gr}}=\left[ e^{x(1+L_1^2/L_2^2+L_1^2/L_3^2+\eps)} -1\right] ^{-1}$.
From here, we see that if $L_1$ is finite and $\eps>-1$ (which is
the case in all regimes if $L_1$ is finite), we will have
$N_{\text{gr}}$ finite, even when $L_2$ and $L_3$ are infinite.
Since, when $L_2$ and $L_3$ are infinite, $N$ is also infinite, we
have $N_{\text{gr}}/N=0$ in all temperature regimes.
This is valid when $L_1$ is finite and also, naturally, when $L_1\tends\infty$. This last scenario can be viewed as an extreme particular case of the anisotropic boxes studied in the thermodynamic limit by van den Berg in \cite{vandenBerg2}.

Inserting the results of \ref{sumsSij} for $S(0,-1)$ in (\ref{eta}), we obtain the asymptotic expansion for $\eta$  	 
	\begin{eqnarray}
    \eta &\sim &
    \frac{\pi^{3/2}}{8}x^{-3/2} \left[
    \z \left( \frac{3}{2}\right) -\z \left( \frac{1}{2}\right)
    \eps x-\frac{1}{8\pi }\z \left( \frac{3}{2}\right)
    (\eps x)^{2}\right] \nonumber\\* &&\mbox{}
    -\frac{\pi}{16}\eps +\frac{\pi}{8}x^{-1}\ln x-
    \frac{\pi}{4}x^{-1}
    \ln \frac{2\sinh (\pi \sqrt {\eps })}{\sqrt{\eps }}  \, .
    \label{etatrunc}
     \end{eqnarray}
Like in the previous two cases, by taking the thermodynamic limit
in this expression we recover the first
order results (\ref{etabulk1}) and (\ref{etabulk2}).

In what concerns the specific heat, the analogue of expressions
(\ref{sh-S3}) and (\ref{sh-1D}) here is
	\begin{equation}
    C=\frac{\pi L_{2}L_{3}}{4L_{1}^{2}}\left[ xS(2,0)+2S(1,-1)+
    \frac{2}{x}S(0,-2)
    -x\frac{\left[ \frac{1}{x}S(0,-1)+S(1,0)\right] ^{2}}{S(0,0)}
    \right]  .
    \label{sh-R2}
	\end{equation}
Hence, using the asymptotic expansions of \ref{sumsSij} yet again yields an asymptotic expansion for the specific heat. However, in order to achieve this, in the specific case of $S(1,-1)$ and $S(0,-2)$ we use (\ref{B:df0at-1v2}) (see last paragraph of \ref{sumsSij}), which is valid only for $|\eps |<1$.

We can now obtain $\eps $ for given $\eta$ from (\ref{etatrunc}) and use it to calculate $\eta _{1}$, which, as we have seen, is the analogue of the three-sphere $N_{\text{gr}}$. The results for $\eta_1/\eta$ are plotted in figure~1 in the cases $\eta=10^2$ and $\eta=10^5$. Like in $S^3$, if fully numerical results were plotted, they would totally superimpose the analytical ones (in fact, the error is even lower in the present case). In figure~2 we plot $C/N$ obtained from (\ref{sh-R2}) together with the expansions for the sums $S(i,j)$. The point where the curves change the pattern in
figure~2 is where $\eps $ becomes larger than 1,
rendering our results for $S(1,-1)$ and $S(0,-2)$ not valid any more and making us resort to fully numerical calculations (when $\eps >1$). In both figures 1 and 2, the curves are similar to those pertaining to the three-sphere. In the large $N$ limit they tend
to be the same, specifically, the standard text book curves of the
infinite box. However, we see that the finite size effects have some differences from one situation to the other. Namely, the condensate fraction approaches the bulk limit curve faster in S$^3$ with increasing $N$ than in the infinite slab with increasing $\eta$ and the specific heat peak is higher and at a higher temperature in the infinite slab than in S$^3$. In the next section, these differences are brought out analytically.

\section{Critical region temperature expansions}
\label{critical}

Expressions (\ref{Ntrunc}) and (\ref{etatrunc}) give us
$\eps $ implicitly as a function of $N$ or $\eta$ and from here we
have all thermodynamic quantities. However, it would be much more
convenient to have $\eps $ given explicitly. That is what we
aim for in this section, concentrating on the critical region. We
solve (\ref{Ntrunc}) and (\ref{etatrunc}) perturbatively for
$\eps$. This enables us to obtain expansions for $\eps$ (and hence, the chemical potential) and specific heat
in powers of $(T/T_0-1)$ (where
$T_0$ is defined below). The effects of finite size and geometry are contained in the coefficients of these expansions.
In this way, we obtain an analytical description of how
the system approaches critical behaviour as the thermodynamic limit is approached.

\subsection{The three-sphere}
\label{critical:S3}

Define $x_{0}$ and respective temperature $T_0$ as being the values of $x$ and $T$ at which the chemical potential assumes the critical value $\mu =0$ (hence, at which $\eps =0$). From
(\ref{Ntrunc}), we have
     \begin{equation}
    N=\frac{\sqrt{\pi }}{4}\z \left( \frac{3}{2}\right)
    x_{0}^{-3/2}-\frac{1}{2}x_{0}^{-1}  .
    \label{x0-S3}
     \end{equation}
Naturally, in the thermodynamic limit we have $x_0/x_c\tends 1$ and
for large particle numbers $x_0$ will be very close to $x_c$. More precisely, $T_c/T_0=x_0/x_c\simeq 1-0.29x_c^{1/2}$. Expanding $\eps$ in powers of $(x/x_0)^{1/2}-1$ around $x=x_0$ we have
 \begin{equation}
\eps =a_1\left[ \left( \frac{x}{x_0}\right)^{1/2} -1\right] +a_2\left[ \left( \frac{x}{x_0}\right)^{1/2} -1\right] ^2+\cdots
\label{epsexpS3}
 \end{equation}
for some coefficients $a_1$ and $a_2$. Using (\ref{Ntrunc}) we find for $a_1$
     \[
    a_{1}=-x_{0}^{-1/2}\frac{(3\sqrt{\pi }/4)
    \z \left( 3/2\right) -x_{0}^{1/2}}{
    \pi ^{2}6+(\sqrt{\pi }/4)\z \left( 1/2
    \right) x_{0}^{1/2}} \, .
     \]
For large $N$, $x_{0}\ll 1$ and we have to the leading order,
     \begin{equation}
    a_{1}\simeq -\frac{9\z \left( 3/2\right) }{2\pi
    ^{3/2}}x_{0}^{-1/2}  .
    \label{a1S3}
     \end{equation}
The result for $a_{2}$ is quite long and there is no point in
displaying it in full. For $x_{0}\ll 1$ it becomes, to leading
order,
     \begin{equation}
    a_{2}\simeq \frac{27\z \left( 3/2\right) ^{2}}{20\pi }
    x_{0}^{-1}  .
    \label{a2S3}
     \end{equation}
Inserting (\ref{a1S3}) and (\ref{a2S3}) in
(\ref{epsexpS3}), we finally have the approximation
	\begin{equation}
    \eps\simeq -2.11x_0^{-1/2}\left[ \left( \frac{x}{x_0}\right)^{1/2} -1\right] 
    +2.93x_0^{-1}
    \left[ \left( \frac{x}{x_0}\right)^{1/2} -1\right]^2
    +\cdots.
    \label{epsdelta}
	\end{equation}
To make this
expansion more readily interpreted, it can be put in terms of number of particles and temperature if we use (\ref{x0-S3}) and expand $\sqrt{x/x_0}=\sqrt{T_0/T}$ in powers of $(T/T_0-1)$, yielding
     \begin{equation}
    \eps\simeq 1.01N^{1/3}\left( \frac{T}{T_0} -1\right)+0.66N^{2/3}
    \left( \frac{T}{T_0} -1\right)^2
    +\cdots.
    \label{epst-S3}
     \end{equation}
Hence, in a neighborhood of $T_0$, we have $\eps$ (or the chemical potential, if we use (\ref{x,eps})) given explicitly as a function of the temperature, for a given number of
particles. In this expression, we clearly see $\eps$ becoming
increasingly steeper at $T_0$ as the thermodynamic limit is
approached ($N\tends \infty$), a fact that could be an\-ti\-ci\-pa\-ted from the bulk expressions. Because of
the way in which we expanded the hyperbolic cotangent in (\ref{Ntrunc}), these
results are valid only for $|\eps|<1$ (and of course, for small
$T/T_0-1$). As can be seen from (\ref{epst-S3}), this is satisfied
if $N^{1/3}(T/T_0-1)$ is sufficiently small. Therefore, the
temperature window for which (\ref{epst-S3}) is valid is
increasingly narrow as we approach the thermodynamic limit.

We can now find an expansion for the specific heat by inserting
(\ref{epsexpS3}) in (\ref{A:asympt1}), which in turn is inserted in (\ref{sh-S3}). We obtain
     \begin{eqnarray}
    \frac{C}{N}&\simeq &
    \frac{15\z
    \left( \frac{5}{2}\right) }{4
    \z \left( \frac{3}{2}\right) }+\left[
    \frac{81\z \left( \frac{3}{2}\right) ^{2}}{40\pi }-
    \frac{45\z \left( \frac{5}{2}\right) }{4
    \z \left( \frac{3}{2}\right) }\right] \left[ \left( \frac{x}{x_0}\right)^{1/2} -1\right] 
    \nonumber \\*   && \mbox{}
    -\frac{729\z \left( \frac{3}{2}\right) ^{3}}{1400\sqrt{\pi }}
    x_{0}^{-1/2}\left[ \left( \frac{x}{x_0}\right)^{1/2} -1\right] ^{2}+\cdots \nonumber \\ 
      &\simeq  &
    1.93+0.69\left(  \frac{T}{T_0} -1\right) -1.25N^{1/3}
    \left(  \frac{T}{T_0} -1\right)^{2} +\cdots. 
    \label{shexp-S3}
     \end{eqnarray}
The coefficient of each term is given only to highest order in
$x_{0}^{-1/2}$ (or $N^{1/3}$). The first term is the familiar specific heat maximum
of three-dimensional infinite space. The other terms
are specific to the three-sphere. From the coefficients in
(\ref{shexp-S3}) we can calculate the first and second derivatives of
the specific heat with respect to $T$ at $T=T_0$. The first derivative does not depend on
$N$, but the second one goes to infinity as $N\rightarrow \infty$. 
This is not surprising; it is the
renowned specific heat peak growing sharper. By differentiating
(\ref{shexp-S3}) with respect to $T$ and equating
the result to zero, we obtain the position of the specific heat
maximum to be $T_{\text{max}}/T_0\simeq 1+0.28N^{-1/3}$, to first
order in the number of particles. Replacing $T_{\text{max}}$ in (\ref{epst-S3}), we obtain
roughly $\eps\simeq 0.3$. So, $T_{\text{max}}$ falls within the
region where the approximation is valid. This is confirmed by comparison of our analytical results with those of our numerical calculations.

We also present $C/N$ at $T_{0}$ expanded for
small $x_{0}$ (large $N$):
     \begin{eqnarray}
    	\frac{C}{N}(x_{0})&\simeq &\frac{15\z \left( \frac{5}{2}\right) }{4
    \z \left( \frac{3}{2}\right) }+\left[
    \frac{15\z \left( \frac{5}{2}\right) }{2\sqrt{\pi }
    \z \left( \frac{3}{2}\right) ^{2}}-
    \frac{27\z \left( \frac{3}{2}\right) }{8\pi ^{3/2}}
    \right] x_{0}^{1/2}+\cdots \nonumber \\
    &\simeq &1.93-0.79N^{-1/3}+\cdots ,
    \label{Cexp-S3}
     \end{eqnarray}
a formula that also agrees very well with the numerical results.
The sub-leading term gives us the rate at which the specific heat
at $T_0$ is decreasing as we get away from the thermodynamic
limit.

By inserting (\ref{epsdelta}) or (\ref{epst-S3}) in the expression for the ground state population, we immediately have the condensate fraction as a function of temperature in the critical region. For purposes of comparison with the infinite slab geometry, we limit ourselves to presenting the leading behaviour at $T=T_0$: 
\[
\frac{N_\text{gr}}{N}\simeq \left(  \frac{\sqrt{\pi}\zeta \left( 3/2\right) }{4}\right)^{-2/3}N^{-1/3}\simeq 0.91N^{-1/3}.
\]

\subsection{The infinite slab}
\label{critical:slab}

For the infinite slab a similar analysis to that described in the previous section can be followed. 
In this case, $x_{0}$ is still defined as the value of $x$ at which
$\mu =0$ and is given from (\ref{etatrunc}) as
     \[
    \eta =\frac{\pi^{3/2}}{8}\z \left( \frac{3}{2}\right)
    x_{0}^{-3/2}+\frac{\pi}{8}x_{0}^{-1}\ln x_{0}-
    \frac{\pi}{4}x_{0}^{-1}\ln (2\pi ) \, .
     \]
We now have 
\[
\frac{T_c}{T_0}=\frac{x_0}{x_c}\simeq 1+0.14x_c^{1/2}\ln \frac{x_c}{4\pi^2}\, .
\]
Naturally, as before $\displaystyle{\lim_{N\rightarrow \infty } x_{0}/x_{c}=1}$. Comparing this expression for $T_c/T_0$ with the same one for the three-sphere, we see that in both cases $T_0$ is higher than $T_c$. However, due to the presence of the log factor in the present case, we have that $T_0/T_c$  is higher in this case than in S$^3$. This is consistent with the fact that the specific heat peaks in the infinite slab are to the right of the respective peaks in the three-sphere.

The result for $\eps $ is
     \begin{eqnarray}
    \eps &\simeq &
    -\frac{9\z \left( \frac{3}{2}\right) }{\pi ^{3/2}}
    x_{0}^{-1/2}\left[ \left( \frac{x}{x_0}\right)^{1/2} -1\right] 
    +\frac{27\z \left( \frac{3}{2}\right) ^{2}}{10\pi }
    x_{0}^{-1}\left[ \left( \frac{x}{x_0}\right)^{1/2} -1\right] ^{2}+\cdots \nonumber \\
    &\simeq &1.73\eta^{1/3} \left(  \frac{T}{T_0} -1\right)
    +0.98\eta^{2/3} \left(  \frac{T}{T_0} -1\right)^2
    +\cdots.
    \label{epsexp-R2}
     \end{eqnarray}
The coefficients
are only displayed to highest order in $x_{0}^{-1/2}$. For the specific heat we obtain
     \begin{eqnarray}
    \frac{C}{N}&\simeq &
    \frac{15\z \left( \frac{5}{2}\right) }{4
    \z \left( \frac{3}{2}\right) }+\left[
    \frac{81\z \left( \frac{3}{2}\right) ^{2}}{20\pi }-
    \frac{45\z \left( \frac{5}{2}\right) }{4
    \z \left( \frac{3}{2}\right) }\right] \left[ \left( \frac{x}{x_0}\right)^{1/2} -1\right] 
    \nonumber \\*   &&
    \mbox{}-\frac{243\z \left( \frac{3}{2}\right) ^{3}}{1400\sqrt{\pi }}
    x_{0}^{-1/2}\left[ \left( \frac{x}{x_0}\right)^{1/2} -1\right] ^{2}+\cdots \nonumber \\  
    &\simeq &
    1.93-1.51\left(  \frac{T}{T_0} -1\right) -0.36\eta^{1/3}
    \left(  \frac{T}{T_0} -1\right)^2 +\cdots.
    \label{Cexp-R2}
     \end{eqnarray}
Comparing (\ref{Cexp-R2}) with (\ref{shexp-S3}) we see that
the qualitative behaviour is very similar except that now the
second term has a negative coefficient, meaning that $T_{0}$ is to
the right of the specific heat maximum, in contrast to what happens
in the three-sphere. The second derivative of the specific heat with respect to $T$ still becomes
infinitely negative as $\eta\rightarrow \infty$. Once more, this
corresponds to the sharpening of the specific heat peak.

If we try to get the value, $T_{\text{max}}$, where the specific
heat has its maximum, as we did in S$^3$, we
get $T_{\text{max}}/T_0\simeq 1-2.11\eta^{-1/3}$. However, this result
is not trustworthy because the value $\eta^{1/3}(T_{\text{max}}/T_0-1)
\simeq -2.11$ is quite large, rendering (\ref{epsexp-R2}) not
valid for $T=T_{\text{max}}$.

The expansion of the specific heat at $T_0$ for small $x_0$ is
     \begin{eqnarray}
    \frac{C}{N}(x_{0})&\simeq &\frac{15\z \left( \frac{5}{2}\right) }{4
    \z \left( \frac{3}{2}\right) }-
    \frac{15\z \left( \frac{5}{2}\right) }{4\sqrt{\pi }
    \z \left( \frac{3}{2}\right) ^{2}}x_{0}^{1/2}
    \ln \frac{x_{0}}{(2\pi )^{2}}
    -\left[
    \frac{\pi ^{3/2}}{3\z \left( \frac{3}{2}\right) }+
    \frac{27\z \left( \frac{3}{2}\right) }{4\pi ^{3/2}}
    \right] x_{0}^{1/2}+\cdots \nonumber \\ 
    &\simeq &1.93+0.34\eta^{-1/3}\ln \eta-3.07
    \eta^{-1/3}+\cdots,
    \label{C at x0-R2}
     \end{eqnarray}
which is the analogue of (\ref{Cexp-S3}) for the infinite slab. This expansion converges more slowly than its equivalent in S$^3$. For this reason, the expression in (\ref{C at x0-R2}) is not accurate for low particle numbers like $\eta=10^2$. Naturally, it gains precision as $\eta$ increases or as more terms are included. We see that here the leading finite size correction is positive for large enough particle numbers (specifically, for $\ln x_0\lesssim -5.6$ which corresponds to $\eta \gtrsim 8.6\times 10^3$). This is opposite to what happens in the previous case, where the leading finite size correction is always negative, and consistent with the fact that the specific heat peaks are higher in the infinite slab than in S$^3$.

For the condensate fraction, at the temperature $T_0$ we have to first order  $\eta_1/\eta\simeq (2/3)\zeta\left( 3/2\right)^{-2/3}\eta^{-1/3}\ln \eta\simeq 0.35\eta^{-1/3}\ln\eta$. Comparing with the corresponding expression in the three-sphere, we observe that there is an extra logarithmic factor. This means that in the present case the condensate fraction at $T_0$ does not tend to zero as quickly as in S$^3$ when the thermodynamic limit is approached. Indeed, this can be observed in figure~1.

\section{Conclusion}
\label{conclusion}

We studied the finite size effects on the thermodynamics of an ideal Bose gas at low temperature, especially in the BEC regime. For this, we obtained asymptotic expansions for the number of particles and specific heat sums which contain the bulk behaviour in their leading contributions, while the other contributions contain the finite size corrections. The three cases studied have mathematical similarities that enabled a unified treatment. The asymptotic expansions proved very accurate in the BEC regime and also in the vicinity of the critical point, even when keeping only two or three terms. They are not as useful in the one-dimensional box due to the absence of BEC in this case.

We were able to provide expressions for the chemical potential and specific heat as explicit functions of the temperature only in the vicinity of the critical point,
in the form of expansions whose coefficients depend only on the size and geometry of the system.
These expressions provide an analytical description of
how thermodynamic functions approach their bulk forms as the systems approach the thermodynamic limit, showing for example the gradual sharpening of the specific heat peak. Furthermore, they allow us to see clearly the effects of finite size and the differences between these effects in different geometries. For instance, while in the three-sphere the specific heat maximum happens at a temperature $T>T_0$ the opposite happens in the infinite slab and the leading finite size correction to the specific heat value at $T_0$ is negative in the former case and positive in the latter.

It should be possible to apply the methods employed here to study BEC in other systems. In particular, it is likely that the relativistic ideal Bose gas can be treated in this way. As mentioned in the introduction, finite size effects for relativistic bosons in the three-sphere have been studied by several authors \cite{Trucks,AltaieMalkawi,SinghPathria}
 using a combination of analytical and numerical techniques, but fully analytical results have not yet been obtained for that system to the accuracy that we have described here. 
It would also be interesting to consider the finite size systems studied here with an interacting Bose gas and see to what extent our ideal gas results would be affected.

\appendix
    
\section{The sums $S(i,j)$}
\label{sumsSij}

In this appendix, we consider the class of sums
 \begin{equation}
S(i,j)=\sum_{n=1}^{\infty}\sum_{k=1}^{\infty} n^{2i}k^je^{-kx(n^2+\eps)}
\label{A:def-S}
 \end{equation}
which appear in the expressions for the number of particles and specific heat in the different cases studied. We will obtain asymptotic expansions for small $\eps x$. For our applications, we need to consider the cases $i=0,1,2,3$ and $j=-2,-1,0,1$.

Apply the Mellin-Barnes transform
 \begin{equation}
e^{-x}=\frac{1}{2\pi i}\int_{c-i\infty}^{c+i\infty} d \alpha\, \Gamma(\alpha)x^{-\alpha}  , \quad
    x>0 , \; \, c>0 
    \label{A:MB}
 \end{equation}
to the exponential in (\ref{A:def-S}). Then, we can move the summations inside the integral sign as long as there is uniform convergence, which is the case if we choose $c>\max \{0,j+1,i+1/2\}$. The sum in $k$ becomes a Riemann $\zeta$-function (which we denote by $\zeta$) and we have
 \begin{equation}
S(i,j)=\frac{1}{2\pi i}\int_{c-i\infty}^{c+i\infty} d \alpha\, f_i(\alpha,\eps)\Gamma(\alpha)
\zeta(\alpha -j)x^{-\alpha}  , 
 \quad c>\max \left\{0,j+1,i+\frac{1}{2}\right\}  ,
 \label{A:S-MB}
	\end{equation}
 where we defined the functions $f_i$ by
  \begin{equation}
 f_i(\alpha,\eps)=\sum_{n=1}^{\infty}n^{2i}(n^2+\eps)^{-\alpha}\; ,\quad \Re(\alpha)>i+\frac{1}{2}\, .
 \label{A:def-f}
  \end{equation}
 The functions $f_i$ are dealt with in \ref{functionsfi}. There, it is shown that they can be analytically continued to the whole complex plane except for isolated singularities at $\alpha=1/2+i-k$, $k=0,1,2,\ldots$. The other factors are also analytical in $\mathbb{C}$ except for poles at $\alpha=j+1$ (coming from $\zeta(\alpha-j)$) and at $\alpha=0,-1,-2,\ldots$ (coming from $\Gamma (\alpha)$). We now close the vertical path of integration in the integral in (\ref{A:S-MB}) in a large rectangle to the left and solve it using the residue theorem. For this, we must know the residues of the integrand. In what concerns $f_i(\alpha,\eps)$, all the information required is given in \ref{functionsfi}. The residues of $\Gamma (\alpha)$ at $\alpha=-k$ are $(-1)^k/k!$ and the residue of $\zeta(\alpha)$ at $\alpha=1$ is 1. Due to the decaying properties of the integrand for large values of $\Im (\alpha)$, the expansions thereby obtained are asymptotic for small $|\eps x|$. These are, for $i=1,2,3$ and $j=0,1$,
 \begin{equation}
 S(i,j)\sim f_i(j+1,\eps)x^{-j-1}+
 \frac{1}{2}x^{-i-1/2}\sum_{k=0}^{\infty}\frac{(-1)^k}{k!}\Gamma \left(\frac{1}{2}+i\right) \zeta\left(\frac{1}{2}+i-j-k\right) (\eps x)^k .
 \label{A:asympt1}
  \end{equation}
 For $i=0$ and $j=0,1$, we have
   \begin{eqnarray}
 S(i,j)&\sim &f_0(j+1,\eps)x^{-j-1}
 +\frac{\sqrt{\pi}}{2}x^{-1/2}\sum_{k=0}^{\infty}\frac{(-1)^k}{k!}
 \zeta\left(\frac{1}{2}-j-k\right) (\eps x)^k \nonumber\\
 &&\mbox{}+\frac{1}{2}\sum_{k=0}^{\infty}\frac{(-1)^{k+1}}{k!}
 \zeta\left( -j-k\right) (\eps x)^k  .
 \label{A:asympt2}
  \end{eqnarray}
 The first term in (\ref{A:asympt1}) comes from the $\zeta$-function pole, while the infinite summation comes from the poles of $f_i$. In (\ref{A:asympt2}) it is exactly the same except that we have a second infinite summation. This comes from the $\Gamma$-function poles. It is not present in (\ref{A:asympt1}) because these poles are removable in this case, due to the zeros of $f_i(\alpha,\eps)$ when $i\neq 0$.
 
 When $j=-1,-2$ the situation is only slightly different in that the pole coming from the $\zeta$-function is double and so the term with $k=-j-1$ is omitted from the second summation in (\ref{A:asympt2}) and the first term in (\ref{A:asympt1}) and (\ref{A:asympt2}) will be different. Specifically, it will be given by the residue of $f_i(\alpha,\eps)\Gamma (\alpha)\zeta (\alpha -j)x^{-\alpha}$ at $\alpha=j+1$. This is obtained in a straightforward fashion from the results of \ref{functionsfi}, more precisely, from (\ref{B:recur}), (\ref{B:f0at-k}), (\ref{B:df0at0}) and (\ref{B:df0at-1v2}).

\section{The functions $f_i(\alpha,\eps)$}
\label{functionsfi}

In \ref{sumsSij}, equation (\ref{A:def-f}), we have defined the functions $f_i(\alpha,\eps)$. Here we obtain their meromorphic structure in the variable $\alpha$ as well as simple expressions for $f_i(\alpha,\eps)$ and $(\partial /\partial\alpha)f_i(\alpha,\eps)$ at some particular values of $\alpha$. We need this information for the calculations in \ref{sumsSij}.

First notice that $f_i(\alpha,\eps)$ for $i=1,2,3$ is easily obtained from $f_0(\alpha,\eps)$ by using the recurrence relation
 \begin{equation}
f_{i+1}(\alpha,\eps)=f_i(\alpha -1,\eps)-\eps f_i(\alpha,\eps) .
\label{B:recur}
 \end{equation} 
Hence, we only need to study $f_0(\alpha,\eps)$. Equation (\ref{A:def-f}) defines $f_0(\alpha,\eps)$ as an analytical function of $\alpha$ for $\Re (\alpha)>1/2$. There is more than one way of obtaining the analytical continuation of this function. Ghika and Visinescu \cite{GV}, Ford \cite{Ford80} and Elizalde and Romeo \cite{ER} all found procedures to do it, but only for $\eps >0$ (whereas we have $\eps >-1$). (Ford's result is actually for a more general function of which ours is a particular case.) The same function as in \cite{Ford80} was treated by Actor \cite{Actor}, who obtained the analytical continuation in the case $|\eps|<1$. The procedure followed in \cite{Toms06} for a similar function would yield the required pole structure (in the whole range of $\eps$). Here we show a different route. 

The analytical continuation can be achieved by noting that $f_0(\alpha,0)=\zeta (2\alpha)$ and that
 \begin{equation}
\frac{\partial}{\partial\eps}f_0(\alpha,\eps)=-\alpha f_0(\alpha +1,\eps),
\label{B:relation-dif}
 \end{equation}
or in its integral version,
 \begin{equation}
f_0(\alpha,\eps)=\zeta (2\alpha)-\alpha \int_0^{\eps}f_0(\alpha +1,\eps)\,  d \eps \, .
\label{B:relation-int}
 \end{equation}
This immediately provides the analytical continuation to $\Re (\alpha)>-1/2$. In this region, we see that the only pole will be the one of $\zeta (2\alpha)$ at $\alpha=1/2$, its residue being 1/2. The other poles and residues follow easily by using (\ref{B:relation-int}) repeatedly, replacing $f_0(\alpha,\eps)$ and $f_0(\alpha +1,\eps)$ by their Laurent expansions. Specifically, $f_0(\alpha,\eps)$ is analytical in $\mathbb{C}$ except for simple poles at $\alpha =1/2-k$, $k=0,1,2,\ldots$ and the respective residues are given by
 \[
\text{Res}_k(\eps)=-\left( \frac{1}{2}-k\right) \int_0^{\eps}\text{Res}_{k-1}(\eps)\,  d \eps \, ,\quad k=1,2,\ldots ,
 \]
where $\text{Res}_k(\eps)$ denotes the residue of $f_0(\alpha,\eps)$ at $\alpha=1/2-k$. By induction, we finally obtain
 \[
\text{Res}_k(\eps)=(-1)^k\frac{1}{2}\left(\frac{1}{2}-1\right)\cdots\left(\frac{1}{2}-k\right)\frac{\eps ^k}{k!}
=\frac{\sqrt{\pi}}{2}\frac{(-\eps)^k}{\Gamma\left( 1/2-k\right) k!}  \, .
 \]
From (\ref{B:recur}) it is easily seen that $f_i(\alpha,\eps)$ is analytical in $\mathbb{C}$ except for simple poles at $\alpha =1/2+i-k$, $k=0,1,2,\ldots$

Expressions for  $f_0(\alpha,\eps)$ at $\alpha=2,1,0,-1,\ldots$ are also needed. The case $\alpha=1$ is given by \cite{Gradshteyn}
 \begin{equation}
f_0(1,\eps)=\sum_{n=1}^{\infty}(n^2+\eps)^{-1}
=\frac{\pi }{2\sqrt{\eps }}\coth (\pi \sqrt{\eps})
    -\frac{1}{2\eps } \, .
\label{B:f0at1}
 \end{equation}
$f_0(0,\eps)$ follows directly from (\ref{B:relation-int}) with $\alpha=0$ yielding $f_0(0,\eps)=-1/2$. The other values are obtained from these ones by using (\ref{B:relation-dif}) or (\ref{B:relation-int}) as needed, yielding
 \begin{equation}
f_0(2,\eps )=\frac{\pi ^{2}}{4\eps }\text{csch}^{2}(\pi
    \sqrt{\eps })+\frac{\pi }{4\eps ^{3/2}}\coth (\pi \sqrt{\eps
    })-\frac{1}{2\eps ^{2}} \, ,
    \end{equation}
    \begin{equation}
f_0(-k,\eps)=-\frac{1}{2}\eps ^k ,\quad k=0,1,2,\ldots
\label{B:f0at-k}
\end{equation}

Now, we want the derivatives $\partial f_0/\partial \alpha $ at
$\alpha =0$ and $\alpha =-1$. Inserting the Taylor series of
$f_0(\alpha ,\eps )$ and $f_0(1+\alpha ,\eps )$ around $\alpha=0$ in
(\ref{B:relation-int}), yields
     \begin{equation}
    \left. \frac{\partial }{\partial \alpha }
    f_0(\alpha ,\eps )\right| _{\alpha =0}  = 
    -\ln \frac{2\sinh (\pi \sqrt{\eps })}{\sqrt{\eps }}  ,
    \label{B:df0at0}
     \end{equation}
where we used (\ref{B:f0at1}). Following the same
procedure for $\partial f_0/\partial \alpha $ at $\alpha =-1$, we
obtain
     \begin{equation}
    \left. \frac{\partial }{\partial \alpha }
    f_0(\alpha ,\eps )\right| _{\alpha =-1}=2\z '(-2)+\frac{1}{2}\eps -\int_{0}^{\eps }
    \ln \frac{2\sinh (\pi \sqrt{\eps })}{\sqrt{\eps }}\, d\eps 
    \, ,\label{B:df0at-1v1}
     \end{equation}
where the prime in $\zeta '(-2)$ means derivative.
There is no simple form for the integral in
(\ref{B:df0at-1v1}). However, in the case of $|\eps |<1$, it
can be expressed in terms of an infinite sum (see, e.g., \cite{Gradshteyn}), which is more convenient than just leaving it as it is. We then have
 	\begin{equation}    
    \left. \frac{\partial }{\partial \alpha }
    f_0(\alpha ,\eps )\right|_{\alpha =-1}=-\frac{\z (3)}{2\pi ^{2}}-\eps
    \ln \frac{2\sinh (\pi \sqrt{\eps })}{\sqrt{\eps }}
    +\sum_{k=0}^{\infty }
    \frac{(2\pi )^{2k}B_{2k}}{(2+2k)(2k)!}\eps ^{k+1}
    \label{B:df0at-1v2}
	\end{equation}
(valid for $|\eps |<1$), where $B_{2k}$ are the Bernoulli numbers.

\begin{figure}[p]
        \includegraphics{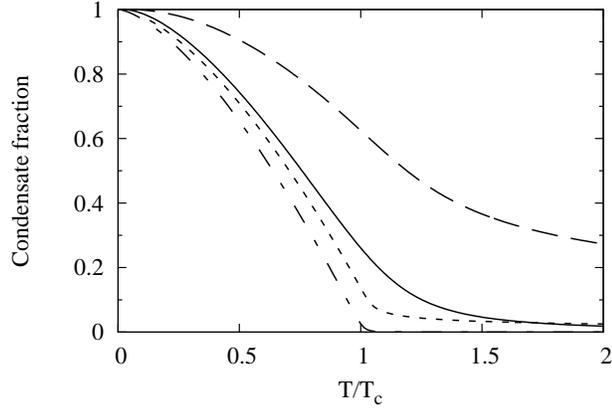}
    \caption{\label{fig1}Condensate fraction $N_{\text{gr}}/N$ in S$^3$ for $N=10^{2}$ (continuous line) and $N=10^{5}$ (alternating short/long-dashed line) and
    $\eta _{1}/\eta $ in the infinite slab for $\eta
    =10^{2}$ (long-dashed line) and $\eta =10^{5}$ (short-dashed line) as
    functions of the rescaled temperature.}
\end{figure}

\begin{figure}
       \includegraphics{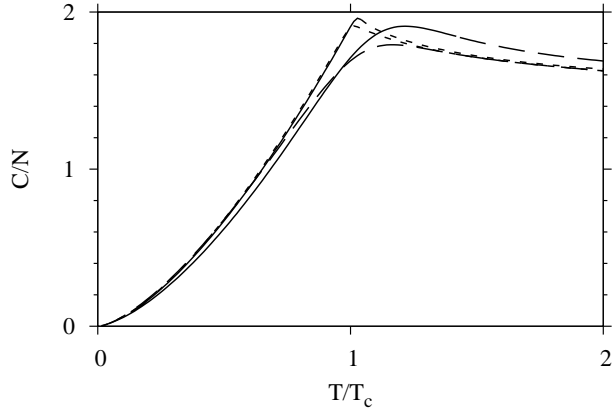}
    \caption{\label{fig2}The specific heat per particle in S$^3$ for $N=10^{2}$ (long-dashed line) and
    $N=10^{5}$ (short-dashed line) and in the
    infinite slab for $\eta =10^{2}$ (continuous
    -analytical- and then long-dashed -numerical- line) and $\eta
    =10^{5}$ (continuous -analytical- and then short-dashed -numerical-
    line) - see text for details.}
\end{figure}

\end{document}